\documentclass[a4paper,english,11pt]{article}
\newcommand{\de}{\delta}

\newcommand{\ka}{\kappa}
\newcommand{\la}{\lambda}
\newcommand{\La}{\Lambda}
\newcommand{\Lao}{\Lambda_0}

\newcommand{\vp}{\varphi}

\newcommand{\vecp}{\vec{{p}}}

\newcommand{\Cll}{C^{\Lambda,\Lambda_0}}
\newcommand{\Lll}{L^{\Lambda,\Lambda_0}}
\newcommand{\Llol}{L^{\Lambda_0,\Lambda_0}}
\newcommand{\ccL}{{\cal L}}
\newcommand{\cLll}{{\cal L}^{\Lambda,\Lambda_0}}

\newcommand{\cLlrn}{{\cal L}^{\Lambda,\Lambda_0}_{r,n}}

\newcommand{\cLol}{{\cal L}^{0,\Lambda_0}}
\newcommand{\cLlol}{{\cal L}^{\Lambda_0,\Lambda_0}}

\newcommand{\pa}{\partial}

\newcommand{\qed}{\hfill \rule {1ex}{1ex}\\ }
\newcommand{\eq}{\begin{equation}}
\newcommand{\eqe}{\end{equation}}
\newcommand{\bea}{\begin{eqnarray}}
\newcommand{\eea}{\end{eqnarray}}
\newcounter{saveeqn}

\begin{document}
\message{reelletc.tex (Version 1.0): Befehle zur Darstellung |R  |N, Aufruf=
z.B. \string\bbbr}
%
%
%  Sonderzeichen
\message{reelletc.tex (Version 1.0): Befehle zur Darstellung |R  |N, Aufruf=
z.B. \string\bbbr}
\font \smallescriptscriptfont = cmr5
\font \smallescriptfont       = cmr5 at 7pt
\font \smalletextfont         = cmr5 at 10pt
\font \tensans                = cmss10
\font \fivesans               = cmss10 at 5pt
\font \sixsans                = cmss10 at 6pt
\font \sevensans              = cmss10 at 7pt
\font \ninesans               = cmss10 at 9pt
\newfam\sansfam
\textfont\sansfam=\tensans\scriptfont\sansfam=\sevensans
\scriptscriptfont\sansfam=\fivesans
\def\sans{\fam\sansfam\tensans}
%----------------------------------------------------------
\def\bbbr{{\rm I\!R}} %reelle Zahlen
\def\bbbn{{\rm I\!N}} %natuerliche Zahlen
\def\bbbE{{\rm I\!E}} %Einheitsmatrix by I. Zoller
\def\bbbm{{\rm I\!M}}
\def\bbbh{{\rm I\!H}}
\def\bbbk{{\rm I\!K}}
\def\bbbd{{\rm I\!D}}
\def\bbbp{{\rm I\!P}}
\def\bbbone{{\mathchoice {\rm 1\mskip-4mu l} {\rm 1\mskip-4mu l}
{\rm 1\mskip-4.5mu l} {\rm 1\mskip-5mu l}}}
\def\bbbc{{\mathchoice {\setbox0=\hbox{$\displaystyle\rm C$}\hbox{\hbox
to0pt{\kern0.4\wd0\vrule height0.9\ht0\hss}\box0}}
{\setbox0=\hbox{$\textstyle\rm C$}\hbox{\hbox
to0pt{\kern0.4\wd0\vrule height0.9\ht0\hss}\box0}}
{\setbox0=\hbox{$\scriptstyle\rm C$}\hbox{\hbox
to0pt{\kern0.4\wd0\vrule height0.9\ht0\hss}\box0}}
{\setbox0=\hbox{$\scriptscriptstyle\rm C$}\hbox{\hbox
to0pt{\kern0.4\wd0\vrule height0.9\ht0\hss}\box0}}}}

\def\bbbe{{\mathchoice {\setbox0=\hbox{\smalletextfont e}\hbox{\raise
0.1\ht0\hbox to0pt{\kern0.4\wd0\vrule width0.3pt height0.7\ht0\hss}\box0}}
{\setbox0=\hbox{\smalletextfont e}\hbox{\raise
0.1\ht0\hbox to0pt{\kern0.4\wd0\vrule width0.3pt height0.7\ht0\hss}\box0}}
{\setbox0=\hbox{\smallescriptfont e}\hbox{\raise
0.1\ht0\hbox to0pt{\kern0.5\wd0\vrule width0.2pt height0.7\ht0\hss}\box0}}
{\setbox0=\hbox{\smallescriptscriptfont e}\hbox{\raise
0.1\ht0\hbox to0pt{\kern0.4\wd0\vrule width0.2pt height0.7\ht0\hss}\box0}}}}

\def\bbbq{{\mathchoice {\setbox0=\hbox{$\displaystyle\rm Q$}\hbox{\raise
0.15\ht0\hbox to0pt{\kern0.4\wd0\vrule height0.8\ht0\hss}\box0}}
{\setbox0=\hbox{$\textstyle\rm Q$}\hbox{\raise
0.15\ht0\hbox to0pt{\kern0.4\wd0\vrule height0.8\ht0\hss}\box0}}
{\setbox0=\hbox{$\scriptstyle\rm Q$}\hbox{\raise
0.15\ht0\hbox to0pt{\kern0.4\wd0\vrule height0.7\ht0\hss}\box0}}
{\setbox0=\hbox{$\scriptscriptstyle\rm Q$}\hbox{\raise
0.15\ht0\hbox to0pt{\kern0.4\wd0\vrule height0.7\ht0\hss}\box0}}}}

\def\bbbt{{\mathchoice {\setbox0=\hbox{$\displaystyle\rm
T$}\hbox{\hbox to0pt{\kern0.3\wd0\vrule height0.9\ht0\hss}\box0}}
{\setbox0=\hbox{$\textstyle\rm T$}\hbox{\hbox
to0pt{\kern0.3\wd0\vrule height0.9\ht0\hss}\box0}}
{\setbox0=\hbox{$\scriptstyle\rm T$}\hbox{\hbox
to0pt{\kern0.3\wd0\vrule height0.9\ht0\hss}\box0}}
{\setbox0=\hbox{$\scriptscriptstyle\rm T$}\hbox{\hbox
to0pt{\kern0.3\wd0\vrule height0.9\ht0\hss}\box0}}}}

\def\bbbs{{\mathchoice
{\setbox0=\hbox{$\displaystyle     \rm S$}\hbox{\raise0.5\ht0\hbox
to0pt{\kern0.35\wd0\vrule height0.45\ht0\hss}\hbox
to0pt{\kern0.55\wd0\vrule height0.5\ht0\hss}\box0}}
{\setbox0=\hbox{$\textstyle        \rm S$}\hbox{\raise0.5\ht0\hbox
to0pt{\kern0.35\wd0\vrule height0.45\ht0\hss}\hbox
to0pt{\kern0.55\wd0\vrule height0.5\ht0\hss}\box0}}
{\setbox0=\hbox{$\scriptstyle      \rm S$}\hbox{\raise0.5\ht0\hbox
to0pt{\kern0.35\wd0\vrule height0.45\ht0\hss}\raise0.05\ht0\hbox
to0pt{\kern0.5\wd0\vrule height0.45\ht0\hss}\box0}}
{\setbox0=\hbox{$\scriptscriptstyle\rm S$}\hbox{\raise0.5\ht0\hbox
to0pt{\kern0.4\wd0\vrule height0.45\ht0\hss}\raise0.05\ht0\hbox
to0pt{\kern0.55\wd0\vrule height0.45\ht0\hss}\box0}}}}

\def\bbbz{{\mathchoice {\hbox{$\sans\textstyle Z\kern-0.4em Z$}}
{\hbox{$\sans\textstyle Z\kern-0.4em Z$}}
{\hbox{$\sans\scriptstyle Z\kern-0.3em Z$}}
{\hbox{$\sans\scriptscriptstyle Z\kern-0.2em Z$}}}}
%rapport.tex
%largemom.tex 
\hspace*{\fill} {\it {\large Theodor  Morschheuser zum }}\\
\hspace*{\fill} {\it {\large 100. Geburtstag gewidmet}}\\[2cm]
\begin{center}
{\Large Large Momentum bounds from Flow Equations}
\end{center}

\begin{center}
{\large \baselineskip 20pt
Christoph Kopper\footnote{email~: 
kopper@cpht.polytechnique.fr} 
\\
and
\\[.2cm]
Fr{\'e}d{\'e}ric Meunier\footnote{email~: 
frederic.meunier@polytechnique.org}
}
\end{center}

\centerline{Centre de Physique Th{\'e}orique de l'Ecole Polytechnique}
\centerline{F-91128 Palaiseau, France}
\date{3 october 2001 }

%\maketitle

\begin{abstract}
\normalsize
We analyse the large momentum behaviour of 4-dimensional
massive euclidean $\vp^4$ theory using the flow equations
of Wilson's renormalization group. 
The flow equations
give access to a simple inductive proof of perturbative
renormalizability. By shar\-pening the induction
hypothesis we prove new  and, as it seems, close to optimal bounds
on the large momentum behaviour of the correlation functions.
The bounds are related to what is generally
called Weinberg's theorem.
\end{abstract}
\newpage

%%%%%%%%%%%%%%%%%%%%%%%%%%%%%%%%%%%%%%%%%%%%%%%%%%%%%%%%%%%%%%%
\section{Introduction }
\vskip.4cm\noindent
The high energy or momentum behaviour of correlation functions
in quantum field theory is of immediate physical interest. It
is reflected in the high energy behaviour of measurable quantities 
such as interaction cross sections. It is also related to questions
of theoretical consistency such as unitarity [1]. 
Four dimensional field theories of physical relevance to this day
have been analysed rigourously in truncated form only,
in particular in perturbation theory. 
The main reason for this is related to the fact that physical
quantities calculated within these theories have to be renormalized,
i.e. reparametrized, since when expressed in the original bare 
parameters of the theory they diverge. 
In the framework of perturbation theory renormalization can be carried
out in full rigour. A particularly attractive tool 
for performing the renormalization proof is the  flow
equation of the Wilson renormalization group [2]. The proof is
considerably simplified as compared to the traditional 
Feynman diagram based proofs, and at the
same time the technical question of eliminating infinities is
traced back to the physical problem of analysing the 
renormalization group flow of the
theory. The statement of renormalizability of the theory then 
can be phrased as follows~: 
On fixing the physical structure (i.e. the field and symmetry
content) and on fixing a finite number of relevant 
parameters by physical renormalization conditions
the perturbative correlation functions of the theory are finite.
  
From these remarks it is obvious that the analysis of the large
momentum behaviour of the correlation functions cannot be
performed rigorously before settling the renormalization issue.
Historically Weinberg [3] performed his famous analysis of the high
energy behaviour of euclidean Feynman amplitudes about ten years before the 
achievement of rigorous renormalization theory. His conviction
that the renormalization procedure would not invali\-date his results was
confirmed in the 70ies, in particular through the work of Berg{\`e}re,
 Lam and Zuber [4]. Their result is of the following form~:
For a given Feynman diagram with given euclidean external momenta
$p_1,\ldots, p_n\,$ the associated Feynman amplitude $I(\la):=\,I(\la
p_1,\ldots,\la p_n)\,$ for $\la\,$ large, has the following asymptotic
expansion 
\[
I(\la)=\, \sum_{r=r_{\max}}^{-\infty} \sum_{s=0}^{s_{\max}}
a_{rs}\,\la ^r (\ln\la)^s\ .
\]
The powers of logarithms are related to the number of 
renormalization operations performed on the graph,
whereas the leading Weinberg power $r_{\max}\,$ 
is the maximal scaling dimension of all 
subgraphs which are irrigated by
the flow of large external momenta. 
Berg{\`e}re, de Calan and Malbouisson [4] generalized the previous result 
to  the situation where only a subset of momenta is scaled by $\la\,$.
As regards the technique of proof, it is based on the Zimmermann
forest formula in parametric space together with the Mellin transform.

Our results are related to those of Weinberg and
followers. Since the flow equations do not require cutting up
perturbative amplitudes into Feynman amplitudes, the result is
stated for the full amplitude, and {\sl it depends on the
geometry of the set of external momenta only}. It is written directly in
general form such that the bound can also be read off in situations
where only subsets of momenta grow large.      

We restrict our considerations
to the simplest item of a renormalizable field theory in four
dimensions. The flow equations have  been used to prove renormalizability
of most theories of physical interest, including theories with massless
fields, and also nonabelian gauge theories [6].
The present considerations could then be extended to those theories
to prove strict UV bounds. The method of proof is in accord with the
standard flow equation inductive proofs. It uses 
sharpened induction hypotheses incorporating 
the improvement of UV behaviour when momentum derivatives are applied
to the correlation functions. In closing we note that the flow
equations have been used extensively in recent years beyond the field
of mathematical physics, in theoretical physics and phenomenology. For
a review see [7].

\section{Renormalisation and large momentum bounds 
from the Flow Equations}
\vskip.4cm \noindent
\subsection{The Flow equation framework}
\vskip.2cm \noindent
Renormalization theory based on the Wilson flow equation (FE) 
has been exposed quite often in
the literature [5]. So we will introduce it rather  shortly.
The object studied is the regularized generating functional
$L^{\La,\Lao}$ of connected (free propagator) amputated 
Green functions (CAG).
The upper indices  
$\La$ and $\Lao$ enter through the regularized propagator
\[
C^{\La,\Lao}(p)\,=\, {1 \over  p^2+m^2} 
\{ e^{- {p^2+m^2 \over \Lao^2}} -e^{- {p^2+m^2 \over \La^2}} \}
\]
or its  Fourier transform 
$\,
\hat{C}^{\La,\Lao}(x)= 
\int_p C^{\La,\Lao}(p) \,e^{ipx}\,$,  with $\int_p\,:= \int_{\bbbr^4} 
 {d^4 p \over (2\pi)^4}\,$.
We assume 
$\,
0 \le \La \le \Lao \le \infty
\,$ 
so that the Wilson flow parameter $\La$ takes the role of an 
infrared (IR) cutoff\footnote{Such a cutoff is of course not necessary
in a massive theory. The IR behaviour is only modified for $\La$ above
$m$.}, whereas  $\Lao$ is the ultraviolet (UV)
re\-gularization. The full propagator is recovered for 
$\La=0$ and $\Lao \to \infty\,$.
For the "fields" and their Fourier transforms we write 
$\,
\hat{\vp}(x) = \int_p  \vp(p) \,e^{ipx}\,$, $\,
{\de \over \de\hat{\vp}(x)} =
(2\pi)^4 
\int_p {\de \over \de \vp(p)}\, e^{-ipx}\,$.
For our purposes the fields $\hat{\vp}(x)\,$ may be assumed to live in
the Schwartz space ${\cal S}(\bbbr^4)$.
For finite $\Lao$ and in finite volume the theory can be given
rigorous meaning starting from the functional integral
\eq
e^{-(L^{\La,\Lao}(\hat{\vp})+ I^{\La,\Lao})}
\,=\, 
\int \, d\mu_{\La,\Lao}(\hat{\phi}) \; 
e^{- \Llol(\hat{\phi}\,+\,\hat{\vp})} \ .
\label{funcin}
\eqe
On the rhs of 
(\ref{funcin}) $\,d\mu_{\La,\Lao}(\hat\phi) $ denotes the (translation
invariant) Gaussian measure with covariance $\hat{C}^{\La,\Lao}(x)$.
The functional $\Llol(\hat\vp)$ is the bare action including
counterterms, viewed as a formal power
series in the renormalized coupling $g\,$. Its general form for 
symmetric $\vp_4^4$
theory is 
\[
   \Llol(\hat{\vp}) = {g \over 4!}  \int \! \! d^4 x \, \hat{\vp}^4(x)  
   \; + 
\]  
\eq
   + \int \! \!d^4 x \,\{{1 \over 2} a(\Lao)\hat{\vp}^2(x) +
    {1 \over 2} b(\Lao) \sum_{\mu=0}^3 (\pa_{\mu}\hat{\vp})^2(x) +
    {1 \over 4!}c(\Lao) \hat{\vp}^4(x)\} \, ,
\label{nawi}
\eqe  
the parameters $a(\Lao),\ b(\Lao),\ c(\Lao)$
fulfill
\eq
a(\Lao) =O(g)\,,\quad b(\Lao),\ c(\Lao) =O(g^2)\ .
\label{con}
\eqe
They are directly related to the standard mass, wave function and
coupling constant counterterms. 
On the lhs of 
(\ref{funcin}) there appears the normalization factor 
$\,e^{-I^{\La,\Lao}} $  which is
due to vacuum contributions. It
diverges in infinite volume so that we can take the infinite volume
limit only when it has been eliminated.
We do not make the finite
volume explicit here since it plays no role in the 
sequel. For a more thorough
discussion see [5] (in particular the last reference).

The FE is obtained from (\ref{funcin}) on differentiating 
w.r.t. $\La\,$. It is a differential equation for the functional
$L^{\La,\Lao}\,$~: 
\eq
\partial_{\La}(\Lll + I^{\La,\Lao} )\,=\, 
\label{feq}
\eqe
\[
\,=\,\frac{1}{2}\,
\langle\frac{\delta}{\delta \hat\vp},(\partial_{\La}\hat \Cll)
\frac{\delta}{\delta \hat\vp}\rangle\Lll
\,-\,
\frac{1}{2}\, \langle \frac{\delta}{\delta
  \hat\vp}\Lll,(\partial_{\La}
\hat \Cll)
\frac{\delta}{\delta \hat\vp} \Lll\rangle \ .
\]
By $\langle\ ,\  \rangle$ we denote the standard scalar product in 
$L_2(\bbbr^4, d^4 x)\,$. Changing to momentum space and
expanding in a formal powers series w.r.t. $g\,$ we write
(with slight abuse of no\-tation)
\[
L^{\La,\Lao}(\vp)\,=\,\sum_{r=1}^{\infty} g^r\,L^{\La,\Lao}_{r}(\vp)\,.
\]
From $L^{\La,\Lao}_{r}(\vp)$ we then obtain the CAG of order $r$
in momentum space  as 
\eq
(2 \pi)^{4(n-1)} \de_{\vp(p_1)} \ldots \de_{\vp(p_n)}
L^{\La,\Lao}_r|_{\vp \equiv 0}
=
\de^{(4)} (p_1+\ldots+p_{n})\, {\cLlrn}(p_1,\ldots,p_{n-1})\ ,
\label{cag}
\eqe
where we have written 
$\delta_{\vp(p)}=\delta/\delta\vp(p)$.
Note that by our definitions the free 
two point function is not contained  in $L^{\La,\Lao}_{r}(\vp)\,$.
This means that $\cLll_{0,2}\,$ vanishes. This is
important for the set-up of the inductive scheme, from which we will 
prove renormalizability below.
The FE (\ref{feq}) rewritten in terms of the CAG (\ref{cag})
takes the following form
\[
\pa_{\La} \pa^w \,\cLlrn (p_1,\ldots p_{n-1}) =
{1 \over 2} \int_k (\pa_{\La}\Cll(k))\,\pa ^w 
\cLll_{r,n+2}(k,-k, p_1,\ldots p_{n-1})
\]
\eq
-
\sum_{\begin{array}{c}r_1+r_2=r,\\
w_1+w_2+w_3=w\\
n_1+n_2=n +2\end{array} }{1 \over 2} 
\Biggl[ \pa^{w_1} \cLll_{r_1,n_1}(p_1,\ldots,p_{n_1-1})\,\times
\label{fequ}
\eqe
\[
\times\,(\pa^{w_3}\pa_{\La}\Cll(p'))\,\,
\pa^{w_2} \cLll_{r_2,n_2}(p_{n_1},\ldots,p_{n})\Biggr]_{ssym}\ ,
\]
\[ \mbox{where }\quad
p'= -p_1 -\ldots -p_{n_1-1}\,= \,p_{n_1} +\ldots +p_{n}\ .
\] 
Here we have written (\ref{fequ})  directly in a form
where also momentum derivatives of the  CAG (\ref{cag})
are performed, and we used the shorthand notations
\[
\pa^w:= \prod_{i=1}^{n-1}\prod_{\mu=0}^{3}
({\pa \over \pa p_{i,\mu}})^{w_{i,\mu}}\ \mbox{ with }\
w=(w_{1,0},\ldots,w_{n-1,3}),\ 
\] 
\[
 |w_i|=\sum_{\mu} w_{i,\mu}\,,\   
|w|=\sum |w_{i}|\,,\  w_{i,\mu}\in \bbbn_0\ .
\]
The symbol $ssym$ 
means summation over those permutations of the momenta 
$p_1,\ldots, p_n$, which do not leave invariant the subsets
$\{p_1,\ldots, p_{n_1-1}\}$ and $\{p_{n_1},\ldots,p_{n}\}$.
Note that the CAG are symmetric  in their momentum arguments by definition.
The simple inductive proof of the renormalizability of $\vp_4^4$ theor
[5] gives the following bounds, which serve at the same time 
as induction hypotheses~:\\
\eq
A)\mbox{ Boundedness}\qquad
|\pa^w \cLlrn(\vec{p})| \le\,
\ka^{4-n-|w| }\,{{\cal P}_1}(log {\ka  \over m})\,
{{\cal P}_2}({|\vec{p}| \over \ka})\,,\qquad\ 
\label{propo1}
\eqe
\eq
B) \mbox{ Convergence }\quad
|\pa_{\Lao} \pa^w \cLlrn(\vec{p})| \le\,
{1\over \Lao^3}\, \ka^{6-n-|w| }\,{{\cal P}_3}(log {\Lao  \over m})\,
{{\cal P}_4}({|\vec{p}| \over \ka})\ .
\label{propo2}
\eqe
Here and in the following we set $\ka=\,\La+m\,$ and use
the shorthand
$\vec{p}=(p_1,\ldots,p_{n-1})$ and $|\vec{p}|=\sup
\{|p_1|,\ldots,|p_n|\}$.
The ${\cal P}_i$ denote 
polynomials 
with nonnegative coefficients, which 
depend on $r,n,|w|,m$,
but not on $\vec{p},\,\La,\,\Lao$. 
The degree of $\,{\cal P}_1\,$ 
can be shown to be bounded by $r+1-n/2\,$ for $n\ge 4\,$ and
by $r-1\,$ for $n=2\,$. 
The statement (\ref{propo2}) implies renormalizability, 
since it proves the limits
$\,\lim_{\Lao \to \infty,\;\La \to 0}  \cLlrn(\vec{p})\,$ to exist
to all orders $r\,$.
But the statement (\ref{propo1}) has to be obtained first
to prove (\ref{propo2}). 
\subsection{Renormalisation together with large momentum bounds}
\vskip.2cm \noindent
The inductive scheme used to prove
(\ref{propo1},\ref{propo2}) will also be used to obtain the new
bounds. What we need is a sharpened induction hypothesis, 
and better control of the high energy improvement
generated by derivatives acting on the Green functions.
We denote by  $\,p_1,\ldots,p_n\,$ a set of external momenta with
$p_1+\ldots+p_n=0\,$,and  we introduce   
\eq
\eta_{i,j}^{(n)}(p_1,\ldots,p_n)=\inf\Bigl\{|p_i+\sum_{k\in J}
p_k|\,/\,J\subset\bigl(\{ 1,..., n\}-\{ i,j \}\bigr)\Bigr\}\ .
\label{eta}
\eqe 
\\
Thus  $\eta_{i,j}^{(n)}\,$ is the smallest subsum of external momenta
which contains $p_i\,$ and which does not contain $p_j\,$. 
Our new bounds are then given by\\[.1cm]
{\bf Proposition 1~:} For $\,0\leq \La \leq \Lao\,$, $\ka =\La+m\,$,  
and for $\, n \ge 4\,$ 
\eq
  |\pa^w {\cal L}_{r,n}^{\La,\Lao}(\vec{p})|\leq \ka^{4-n}\, 
  \prod_{{i=1 \atop i\neq j}}^{n} 
{1 \over \bigl(\sup(\ka,\eta_{i,j}^{(n)})\bigr)^{|w_i|}} 
{\cal P}^{|w|}_{r,n}
\bigl(\log(\sup({|\vec{p}| \over \ka},{\ka\over m}))
\bigr)\; ,\\
\label{propn1}
\eqe
\eq
\mbox{for  }n=2:
  |\pa^w {\cal L}_{r,2}^{\Lambda,\Lambda_0}(p)|\leq
            \sup(|p|,\ka)^{2-|w|}
{\cal P}^{|w|}_{r,2}\bigl(\log(\sup({|p|\over \ka},{\ka\over
       m}))\bigr)\, . 
\label{propn2}
 \eqe
Here ${\cal P}^{|w|}_{r,n}$ are (each time they appear possibly new)
polynomials with nonnegative coefficients which 
depend on $r,n,|w|,m\,$, but not on $\vec{p},\,\La,\,\Lao\,$.
They are of degree
\[
\deg {\cal P}^{|w|}_{r,n} \le  \left\{
 \begin{array}{l}
|r-1-n/2|  \qquad \ \textrm{  if $\,n= 2$,  $|w|\geq 3\,$}\\
|r-n/2|  \qquad \qquad\, \textrm{ if $\,n =2$, $|w|\leq 2\,$ 
or if $n =4$, $|w|\geq 1\,$}\\
|r+1-n/2|  \qquad \ \; \textrm{ otherwise.}\\
 \end{array}
  \right.
\]
\\[.1cm]
{\it Proof~:}
We will use the standard inductive scheme which goes up in $r\,$ and
for given  $r\,$ descends in $n\,$, and for given
$r,\, n\,$ descends in $|w|\,$ starting from some arbitrary
$|w|_{\max}\,$. The rhs of the FE is then prior the lhs in the
inductive order, and the bounds can thus be verified 
for suitable boundary conditions on integrating the rhs of the FE over
$\La\,$, using the bounds of the proposition. 
To start the induction note that 
\[
 {\cal L}_{r,n}^{\La,\Lao}\equiv 0 \quad \mbox{for }\ n>2r+2
\]
(as follows from the connectedness).
Terms with $n+|w| \ge
5$ are integrated down from $\Lao$ to $\La$, since for those terms we
have the boundary conditions at $\La =\,\Lao\,$ 
following from (\ref{nawi})
\[
 \pa^w \,\cLlol_{r,n} (p_1,\ldots p_{n-1}) =0\ \mbox{ for}\quad
 n+|w| 
\ge 5\,,
\]
whereas the terms with $n+|w| \le 4$ at the renormalization point - 
which we choose at zero momentum for simplicity -      
are integrated upwards from $0$ to $\,\La$, since they are fixed 
at $\La=\,0\,$ by
($\Lao$-independent)
renormalization conditions, which define the relevant parameters of the
theory.
From symmetry considerations we deduce the absence of 
nonvanishing renormalization constants apart from those appearing in
(\ref{con}). The Schl{\"o}milch or integrated Taylor formula permits us
to move away from the renormalization point, treating first 
$\cLol_{r,4}$ and then the momentum derivatives of
$\cLol_{r,2}\,$, in descending order.\\
Note that $j$ in (\ref{propn1}) is arbitrary,
so the bound arrived at will be in fact 
\[
  |\pa^w {\cal L}_{r,n}^{\La,\Lao}(\vec{p})|\leq \ka^{4-n}\, 
\inf_{j,1\le j\le n}  \prod_{{i=1 \atop i\neq j}}^{n} 
{1 \over\bigl( \sup(\ka,\eta_{i,j}^{(n)})\bigr)^{|w_i|}} 
{\cal P}^{|w|}_{r,n}
\bigl(\log(\sup({|\vec{p}| \over \ka},{\ka\over m}))
\bigr)\ .
\]
We will choose $j=n\,$ since the proof is 
independent of this choice.\\
A) $n+|w|\geq 5$~:\\
A1) $n\geq 4\,$: Integrating the FE (\ref{feq}) w.r.t. the flow
parameter $\ka'$ from $\ka\,$ to $\Lao+m\,$ gives the
following bound for the {\sl first} term on the rhs of the FE 
\[
\int_\ka^{\Lao+m} d\ka' \int d^4 p~ e^{-{{p^2+m^2}\over \La'^2}}
\, \ka'^{4-(n+2)-3}\,\prod_{i=1}^{n-1} {1 \over
      \bigl({\sup(\ka',\eta_{i,n}^{(n+2)})\bigl)^{|w_i|}}}
\]
\[\times\, {\cal P}^{|w|}_{r,n+2}\bigl(\log
(\sup({|\vec{p}| \over \ka'},
{|p| \over \ka'},{\ka' \over m}))\bigl)\qquad (\La'=\ka'-m)
\]
\[
\leq \int_\ka^{\Lao+m} d\ka'\int d^4({p\over \ka'})\,\,e^{-{p^2 \over
    {\La'^2}}}\ka'^{3-n-|w|}\prod_{i=1}^{n-1} {1\over
    \bigl({\sup(1,{\eta_{i,n}^{(n+2)}\over \ka'})}\bigr)^{|w_i|}}
\]
\[
\times {\cal P}^{|w|}_{r,n+2}\bigl(\log
(\sup({|\vec{p}| \over \ka'},{|p| \over \ka'},
{\ka' \over m}))\bigl)
\]
\[
\leq\ka^{4-n} \prod_{i=1}^{n-1} {1 \over
     \bigl( {\sup(\ka,\eta_{i,n}^{(n)})\bigl)^{|w_i|}}}\
 {\cal P}^{|w|}_{r,n+2}\bigl(\log
(\sup({|\vec{p}| \over \ka},
{\ka \over m}))\bigl)\;,
\]
which satisfies the required bound.
Here we used  the important inequality:
\begin{equation}
\int d^4 x\  e^{-x^2}\ {\cal P}(\log |x|)\, \prod_{i=1}^{k} {1 \over
      \sup(1,|x+a_i|)}\,\leq\,c(k)\,\prod_{i=1}^{k} {1\over
      \sup(1,|a_i|)}
\label{ineq}
\end{equation}
for suitable $c(k)>0\,$. This inequality will again  be used in
the subsequent considerations. It is easily established using the
rapid fall-off of $e ^{-x^2}\,$.

The required bound on the {\sl second} contribution from the rhs of the FE 
(\ref{fequ}) is established when using the 
induction hypothesis for the terms  $\pa^{w_1} {\cal
  L}_{r_1,n_1}^{\La,\Lao}\,$ and $\pa^{w_2} {\cal
  L}_{r_2,n_2}^{\La,\Lao}\,$. The only new ingredient 
needed is a bound
for the derivatives of the regularization factor appearing in this second
term: 
\[
|\partial^w \,e^{-\frac{q^2+m^2}{\La^2}}| \,\leq \,
c(|w|)\; \ka^{-|w|} \; e^{-\frac{q^2}{\La^2}}
\]
for suitable $c(|w|)>0\,$.
Note also that by the induction hypothesis
\[
\deg {\cal P}_{r_1,n_1}^{|w_1|}\,+\,
\deg {\cal P}_{r_2,n_2}^{|w_2|}\,\le\,
 \left\{
 \begin{array}{l}
|r+1-n/2|  \quad \ \textrm{  if $\,n = 4$,  $w=0\,$ or if $\,n \ge 6$ }\\
|r-n/2|  \quad \qquad\, \textrm{ if $\,n =4$, $|w|\geq 1\,$}\\
 \end{array}
  \right.
\]
in all cases (also if $\,|w|\ge 1\,$, and $\,w_1,\,w_2=\,0\,$).\\
A2) The case ($n=2$,$w=3$) which is simpler due to the appearance of
one external momentum only is treated analogously.\\[.1cm]
B) $n+|w|\leq 4$~:\\
For the relevant terms of dimension $\leq 4$ the induction
hypothesis is easily verified at zero
momentum where it agrees with the results from [5] 
\footnote{ We note that when performing the integration
  over $\ka\,$ from $m$  to $\La+m\,$ for the terms with  $n+|w|=4\,$
there appears a logarithm, which is the origin of the polynomial
${\cal P}^{|w|}_{r,n}$, present also at zero momentum.}.  
To extend it to general momenta we shall choose a suitable
integration path from zero to the momentum configuration considered.\\
B1) For $n=2\,$ we proceed in descending order of $|w|\,$
starting from $|w|_{\max}\,$. We use
\[
\pa^{w} \cLll_{r,2}(p)=\,
\pa^{w} \cLll_{r,2}(0)\,+\,
|\sum_\mu p_{\mu} \int_0^1 d\la \, \pa_{\mu}
\pa^{w} \cLll_{r,2}(\la p)| 
\]
and bound the second term with the aid of the induction hypothesis by
\[
|\sum_\mu p_{\mu} \int_0^1 d\la \, \pa_{\mu}
\pa^{w} \cLll_{r,2}(\la p)|\leq \,
\]
\[ 
|p| \int_0^1 { d\la \over 
\bigl(\sup(\la |p|,\ka)\bigr)^{|w|-1}}\,
{\cal P}^{|w|+1}_{r,2}
\bigl(\log(\sup({|p| \over \ka} ,{\ka \over m}))\bigl)\le  
\]
\[
|p| \Bigl(\int_0^{\inf(1,{\ka \over |p|})}
{ d\la \over \ka^{|w|-1}}\,+\,\int_{\inf(1,{\ka \over |p|})}^1
{ d\la \over \,(\la |p|)^{|w|-1}} \Bigl)\,
{\cal P}^{|w|+1}_{r,2}
\bigl(\log(\sup({|p| \over \ka} ,{\ka \over m}))\bigl) 
\]
\[
\leq |p|^{2-|w|}{\cal P}^{|w|}_{r,2}\bigl(\log(\sup({|p| \over \ka },
{\ka \over m}))\bigl)\ .
\]
B2) To prove the proposition for ($n=4,w=0\, $) we will use repeatedly
the \\[.1cm]
{\it Lemma:} For  $\la \in [0,1]$ and $x,\,y \in \bbbr^d\,$, 
if $|x+y|\geq |x|$ then $|\la x+y|\geq
  \la |x|$.\\[.1cm]
{\it Proof:} 
$\ |\la x+y|\geq|x+y|-|(1-\la)x|\geq|x|-(1-\la)|x|=\la
|x|\,$. 
\qed\\[.1cm]
In fact, the case $n=4,w=0\, $ will be treated 
by distinguishing four  different
situations as regards the momentum configurations.
We use the previously established  bounds for the case $n=4,w=1$. 
These bounds
are in terms of the functions $\eta_{i,j}^{(4)}\,$ from
(\ref{eta}). Assuming  (without loss of generality) 
\[
|p_4| \ge |p_1|\,,\ |p_2|\,,\ |p_3|
\]
we realize that $\eta_{i,4}^{(4)}\,$ can
always be realized by a sum of {\it at most two} momenta from the
set $\{p_1\,,\ p_2\,,\ p_3\}\,$. It is then obvious that the
subsequent cases ii) and iv) cover all possible situations.
The cases i) and iii) correspond to exceptional configurations
for which the bound has to be established before proceeding to the
general ones. The four cases are\\
i) $\{p_1\,,\ p_2\,,\ p_3\} =\{0\,,\ q\,,\ v\}$\\ 
ii) $\{p_1\,,\ p_2\,,\ p_3\}$ such that $\inf_i
\eta_{i,4}^{(4)}\,=\,\inf_i  |p_i|$ \\
iii) $\{p_1\,,\ p_2\,,\ p_3\} =\{p\,,-p \,,\ v\}$\\
iv) $\{p_1\,,\ p_2\,,\ p_3\}$ such that $\inf_i \eta_{i,4}^{(4)}\,=\,
\inf_{j\neq k}|p_j+p_k|\,$.\\[.1cm]
i) To prove the proposition in this case we 
use an integrated Taylor formula~: 
\[
|{\cal L}_{r,4}(0,q,v)|\leq
\]
\[
|{\cal L}_{r,4}(0,0,0)|+\sum_{\mu}\int_0^1d\la\,\Bigl(
|q_{\mu}\pa_{q_{\mu}}{\cal L}_{r,4}
(0,\la q,\la v)|+\,|v_{\mu} \pa_{v_{\mu}}{\cal L}_{r,4}
(0,\la q,\la v)|\Bigr)\ .
\]
The second term is bounded using the induction hypothesis: 
\eq
\sum_{i=2,3}|p_i|
\int_0^1d\la\,{ 1\over  \sup(\ka, \eta_{i,4}^{(4)}(\la)\,)}
 {\cal P}^1_{r,4}\bigl(\log
(\sup({|p_4| \over \ka},\,{\ka \over m}))\bigl)\; .
\label{2term}
\eqe
We have written $\eta(\la)\,$ for the $\eta$-parameter in terms of the
scaled variables $\,p_2^{\la}\,=\,\la q\, ,\quad p_3^{\la}\,=\,\la v\,$.
We directly find $\, \eta_{2,4}^{(4)}(\la)\,=\,\la |q|\,,\quad
 \eta_{3,4}^{(4)}(\la)\,=\,\la |v|\,$
and thus obtain the following bound for (\ref{2term})
\[
|q|\,\Bigl(
\int_0^{\inf(1,{\ka \over |q|})}
\!{d\la  \over \ka}\,+\,
\int_{\inf(1,{\ka \over |q|})}^1
\!{d\la \over \la |q|}\, \Bigr)
 {\cal P}^1_{r,4}\bigl(\log
(\sup({|p_4| \over \ka},\,{\ka \over m}))\bigl)\,+\,
\biggl(q\to \,v\, \biggr)
\]
\[
\le
\biggl({|q|\over \ka}{\ka \over |q|}\,+\,\log({|q|+\ka \over \ka})\,+\,
{|v|\over \ka}{\ka \over |v|}\,+\,\log({|v|+\ka \over \ka}) \biggr)\,
 {\cal P}^1_{r,4}\bigl(\log
(\sup({|p_4| \over \ka},\,{\ka \over m}))\bigl)
\]
\[
\le \, 
{\cal P}^0_{r,4}\bigl(\log
(\sup({|p_4| \over \ka},\,{\ka \over m}))\bigl)\,,
\]
which ends the proof of case i).\\
ii)  We assume without loss of generality
$ \inf_i \eta_{i,4}^{(4)}\,=\,|p_1|\,$.
We use again an integrated Taylor formula along the integration path 
$(p_1^{\la},\,p_2^{\la},\,p_3^{\la}\,)=\,
(\la\,p_1,\,p_2,\,p_3+(1-\la)\,p_1\,)\,$.
By the Lemma  we find $\,
\eta_{1,4}^{(4)}(\la) =|p_1^{\la}|=\, \la |p_1|$, $ 
\eta_{3,4}^{(4)}(\la)\ge \la |p_1|\,$.
The boundary term for $\la=\,0\,$ is bounded  through i).
For the second term we bound
\[
|\sum_{\mu}\int_0^1d\la\,\Bigl(
p_{1,\mu}\,\bigl(\pa_{p_{1,\mu}}-\,\pa_{p_{3,\mu}}\bigr)
{\cal L}(p_1^{\la},\,p_2^{\la},\,p_3^{\la}\,)\Bigl)|
\]
\[
\le
|p_{1}|\,\int_0^1d\la\,
\bigl( { 1\over  \sup(\ka, \eta_{1,4}^{(4)}(\la)\,)}\,+\,
{ 1\over  \sup(\ka, \eta_{3,4}^{(4)}(\la)\,)}\bigr)
 {\cal P}^1_{r,4}\bigl(\log
(\sup({|p_4| \over \ka},\,{\ka \over m}))\bigl)
\]
\[
\le
|p_{1}|\,\biggl(\int_0^{\inf(1,{\ka \over |p_1|})} 
{ d\la \over \ka}\,+\,
 \int_{\inf(1,{\ka \over |p_1|})}^1  
{ d\la\over \la |p_1|}\biggr)\, {\cal P}^1_{r,4}\bigl(\log
(\sup({|p_4| \over \ka},\,{\ka \over m}))\bigr)\; ,
\]
which gives the required bound similarly as in i).\\
iii)  We choose the integration path 
$(p_1^{\la},\,p_2^{\la},\,p_3^{\la}\,)=\,
(\la\,p,\,-p,\,v\,)$.
Here we assume without restriction that 
$\,|v| \le |v-(1-\la)p|\,$, otherwise we interchange the role of
$v\,$ and $-v\,$.
The boundary term leads again back to i).
The integral $\int_0^{1} d\la\,$ of the second term
is cut into four pieces
\[
\int_0^{1}\,=\,
\int_0^{\inf(1/2,{\ka \over |p_1|})}\,+\,
\int_{\inf(1/2,{\ka \over |p_1|})}^{1/2}\,+\,
\int_{1/2}^{\sup(1/2,1-{\ka \over |p_1|})} \,+\,
\int_{\sup(1/2,1-{\ka \over |p_1|})}^1\ .
\]
They are bounded in analogy with ii) using 
$\,\eta_{1,4}^{(4)}(\la\,) =\, \la |p_1|\,$ for $\,\la \le 1/2\,$,
$\,\eta_{1,4}^{(4)}(\la\,) =\,(1- \la) |p_1|\,$ for 
$\,\la \ge 1/2\,$, relations easily established with the aid
of the Lemma.\\
iv) We assume without loss of generality 
$\inf_i \eta_{i,4}^{(4)}\,=\,|p_1+p_2|\,$ and integrate along
$(p_1^{\la},\,p_2^{\la},\,p_3^{\la}\,)=\,
(p_1,\,-p_1+\la(p_1+p_2),\,p_3\,)$.
The boundary term has been bounded in iii).
Using the Lemma  again we find
$\,\inf \eta_{2,4}^{(4)}(\la)=\,\la |p_1+p_2|\,$,
and the integration term is then bounded through
\[
|\sum_{\mu}\int_0^1d\la\,\Bigl(
(p_{1,\mu}+p_{2,\mu}) \,\pa_{p_{2,\mu}}
{\cal L}(p_1^{\la},\,p_2^{\la},\,p_3^{\la}\,)\Bigl)|\le\,
\]
\[
|p_{1}+p_{2}|
\biggl(\int_0^{\inf(1,{\ka\over |p_1+p_2|})}\! \!{d\la \over \ka}+\,
 \int_{\inf(1,{\ka\over |p_1+p_2|})}^1 \!  
{d\la \over \la |p_1+p_2|}\biggr)
{\cal P}^1_{r,4}\bigl(\log
(\sup({|p_4^{\la}| \over \ka},\,{\ka \over m}))\bigr)
\]
which gives the required bound as before.\qed\\[.1cm]
Bounds like those of Proposition 1 can also be proven
using regularizations different from the one applied here.
To analyse properties of Green functions in Minkowski space
it is useful to have regulators which stay bounded for large momenta 
in the whole complex plane. 
An example is
\[
C^{\La,\Lao}(p)=\,{1\over p^2+m^2}\,
\Bigl( ({\Lao^2 \over p^2+m^2+\Lao^2 })^k \,-\,
({\La^2 \over p^2+m^2+\La^2 })^k \Bigl)\ .
\]
One realizes that an inequality analogous to (\ref{ineq}) in this case 
requires that $2k > |w|_{\max}+2\,$. Since 
$|w|_{\max}\,$ should be at least 3 (to be able perform the renormalization 
proof for  the two point function), we need $k\ge 3\,$. Then the proof
can be performed as before.    

\subsection{Weighted trees and large momentum fall-off}
\vskip.2cm\noindent
In this section we want to show that for $n\ge 6\,$ the $n$-point functions 
of symmetric massive $\vp_4^4\,$ 
fall off for large external momenta. The following
definitions are required for a precise formulation
of these fall-off properties.
  
A 4-tree of order $r\,$ is defined to be a connected graph without loops
and with a set of $r\ge 1\,$ vertices of coordination number 4. 
The tree has $n\,$ external
lines with $n=\,2r+2$, which are assumed to be 
numbered,
and it has a set ${\cal I}\, $  of internal lines with $|{\cal I}|=r-1\,$. 
We then denote by 
${\cal T}^{4, n}\,$ the set of all 4-trees with $n\,$ external lines.
A weighted 4-tree is a 4-tree with
a weight $\mu(I)=\,2\,$ attached to each $I\in \cal I\,$.
We now define for $1 \le k\le \,n-4\,$ {\it k-times reduced} (weighted) trees
obtained from (weighted) 4-trees:\\
A 0-times reduced tree is a 4-tree.\\
A $\,k$-times reduced tree $T^{(k)}\,$ is obtained from a
$(k-1)$-times reduced tree  $T^{(k-1)}\,$
through the following process:\\
i) by suppressing {\it one} external line of $T^{(k-1)}\,$,\\
ii) by diminishing by {\it one} unit the weight of one 
among those internal lines
of $T^{(k-1)}\,$, which are adjacent to the vertex where the external line was
suppressed (there are at least 1 and at most 3 lines of this type),\\ 
iii) by suppressing any internal line $I\,$ from the tree if it has
acquired $\mu(I)=\,0\,$ through this process, and fusing the two
adjacent vertices into one,\\
iv) by suppressing the vertex from which the external line 
has been removed, in case this vertex has acquired coordination 
number 2 through this
removal. If two internal lines have been attached to this vertex,
they are fused into  a single one and their weights are added. 
If one internal line  had been attached to this vertex,
it had necessarily weight 0 and was removed through iii).\\[.1cm]
It is then easy to realize that a k-times reduced tree $T^{(k)}\,$
with  $n\,$ external lines has the following
properties~:\\ 
a) It is a tree.\\
b) Its vertices have coordination numbers 3 or 4.\\
c) The  weight $\mu(I)\,$ attached to each internal
line $I\in{\cal I}\, $ of $T^{(k)}\,$ satisfies\\  
i) $\mu(I)\in \{1,\,2\}\,,\quad $ 
ii) $\sum_{I\in{\cal I}}\mu(I)=n-4\,$.\\[.1cm]
The set of weighted reduced trees with $n\,$ external lines is denoted by 
${\cal T}^{n,\mu}\,$. 
We will use these trees to bound the lhs of the FE in terms of the
rhs.

To the external lines of a tree $T^{n,\mu}\in {\cal T}^{n,\mu}\,$ 
we associate $n\,$ external incoming momenta
$\vec p=\,(p_1,\ldots,p_n\,)\,$ and write 
$T^{n,\mu}(\vec p)\,$ for the thus 
assigned tree.
Let then $p(I)\,$ be the (uniquely fixed, by momentum conservation)
momentum flowing through the internal line $I\in \cal I\,$.
For given $\ka\,$ the weight factor of an (assigned  weighted) tree
$T^{n,\mu}(\vecp)\,$  (shortly $T\,$) is defined as
\[
g^{\ka}( T)=\,
\prod_{I\in{\cal I}(T)}{1 \over (\sup(\ka,p(I))^{\mu(I)}}\ .
\]
Our statement on the fall-off of the $n$-point-functions is then
the following\\[.1cm]
{\bf Proposition 2}:  For $n \geq 4$ (and with $\ka =\,\La+m\,$)\\
\eq
|\ccL_{r,n}^{\La,\Lao}({\vecp})|
\leq
\sup_{T\in{\cal
T}^{n,\mu}({\vecp})}\,g^{\ka}(T)\
{\cal P}_{r,n}\bigl(\log(\sup({|\vec p| \over \ka},{\ka \over m}))\bigr)\,,
\label{prop2}
\eqe
where $\,\deg{\cal P}_{r,n}\le\,\,r+1-n/2\,$.\\[.1cm]
{\sl Remark:} We could prove without hardly any change a slightly
sharper version of Proposition 2, by restricting the $\sup\,$ 
in (\ref{prop2}) to  $2k\,$-times reduced trees with $k=\,r+1-n/2\,$. 
For $k\,$ sufficiently large, both sets of trees become equal,
however. \\[.1cm]
{\it Proof:}
We again apply  the standard inductive scheme.
In starting we note that the $\ccL_{r,n}\,$ vanish for $n>2r+2\,$ and
are given by a sum over 4-tree graphs for $n=2r+2\,$, which
obviously satisfy the bounds of the proposition.
We also note that for $n=4$ Proposition 2 follows from Proposition 1. 
Thus we assume $n\ge 6\,$.
\\
i) We bound the {\sl first}  term on the rhs of the FE (\ref{fequ}), 
integrated over $\ka\,$:  
\[
\int_\ka^{\Lao+m} \!\!{d\ka' \over\ka'^{3}}
\int d^4 p \; e^{-{{p^2+m^2}\over \La'^2}}
\!\!\!\sup_{T\in{\cal
T}^{n+2,\mu}({\vecp,p,-p})}\!\!\! g^{\ka'}( T)\;
{\cal P}_{r,n+2}\bigl(\log
(\sup({|\vec{p}| \over \ka'},
{|p| \over \ka'},{\ka' \over m}))\bigl)
\]
\[
\leq \int_\ka^{\Lao+m} d\ka'(\ka')^{4-n-1}\,
\int d^4({p\over \ka'})\,\,e^{-{p^2 \over
    {\ka'^2}}}\,
\prod_{I\in {\cal I}(T_{\max}^{\ka'})}
{1 \over \bigl(\sup(1,{|p(I)|\over \ka'})\bigr)^{\mu(I)}}
\]
\[
\times {\cal P}_{r,n+2}\bigl(\log
(\sup({|\vec{p}| \over \ka'},
{|p| \over \ka'},{\ka' \over m}))\bigl)
\]
\[
=\, \int_\ka^{\Lao+m} { d\ka' \over \ka'}\,
\prod_{i=1,2} {1 \over \bigl(\sup(1,{|\hat{p}(I_i)| \over \ka'})\bigl)}\
g^{\ka'}(T_2^{\ka'})\,
{\cal P}_{r,n+2}\bigl(\log
(\sup({|\vec{p}| \over \ka'},
{\ka' \over m}))\bigl)
\]
\eq
\leq \int_\ka^{\Lao+m} {d\ka' \over \ka'}\,
g^{\ka'}(T_2^{\ka'})\,
{\cal P}_{r,n+2}\bigl(\log
(\sup({|\vec{p}| \over \ka'},
{\ka' \over m}))\bigl)
\label{nnn}
\eqe
with the following explanations:
The integral over $p/\ka'\,$ was bounded with the aid 
of the inequality (\ref{ineq}). By $T_{\max}^{\ka'}\,$ we denote a tree
$T\in {\cal T}^{n+2,\mu}(\vecp, p,-p)\,$ of maximal weight for 
given $\ka'\,$. Then we denote by $\,T_2^{\ka'}(\vec p)\,$ or shortly 
$\,T_2^{\ka'}\,$ a twice reduced tree of $T_{\max}^{\ka'}\,$,
obtained by suppressing the two external lines from
$T_{\max}^{\ka'}\,$, which carried the momenta $p,\,-p\,$, and by
diminishing the weight of two internal lines $I_1,\,I_2\,$, 
adjacent to the respective vertices by one unit (it may happen 
that the two vertices and/or lines are identical). And we set
$\,\hat{p}(I_i):=p(I_i)|_{p,-p=0}\,$.
Now we note that 
\eq
\int_\ka^{\Lao+m}\!   {d\ka' \over \ka'}\,
{ g^{\ka'}(T^{\ka'}) \over g^{\ka}(T^{\ka})}\ 
{\cal P}_{r,n+2}\bigl(\log
(\sup({|\vec{p}| \over \ka'},
{\ka' \over m}))\bigl)\,\le
\label{bd}
\eqe
\[
 \le
\log({|\vec p| \over \ka})\ 
\tilde{{\cal P}}_{r,n+2}\bigl(\log
(\sup({|\vec{p}| \over \ka},
{\ka \over m}))\bigl)
\]
and thus obtain finally the required bound for (\ref{nnn})
\[
(\ref{nnn})\,\le \,\sup_{T\in{\cal
T}^{n,\mu}({\vecp})}\,g^{\ka}(T)\
{\cal P}_{r,n}\bigl(\log(\sup({|\vec{p}| \over \ka},
{\ka \over m}))\bigl)\ .
\]
\\[.1cm]
ii) To bound the {\sl second} term on the rhs of (\ref{fequ}) 
we use the inequality $\,
{\ka}^{-3}\ \exp({-{p'}^2 \over {\La}^2})\,\leq\,
\bigl( \sup(\ka,|p'|)\bigl)^{-2}\ {\ka}^{-1}\,
\,$
to obtain straightforwardly the following bound for any given term 
(with $n_1,\ n_2 \ge 4\,$)\footnote{If e.g. $n_1=2\,$ for the first term, 
we use the bound (\ref{propn2}) and  then
\[
{{\ka}^{-1} \over \bigl(\sup(\ka,|p|)\bigl)^2}\,
 \bigl( \sup(|p|,\ka)\bigr)^{2}
\,{\cal P}_{r,2}\bigl(\log(\sup({|p|\over \ka},{\ka\over m}))\bigr) 
\,\le \, 
{\ka}^{-1} \,
{\cal P}_{r,2}\bigl(\log(\sup({|p|\over \ka},{\ka\over m}))\bigr)\,, 
\]
and retain the contribution of the second term to verify the bound
as in (\ref{snd},\ref{sndd}).}
in the sum appearing on the rhs
of (\ref{fequ}): 
\eq
{{\ka}^{-1} \over \bigl(\sup(\ka,|p'|)\bigl)^2}\,
\sup_{T_1\in {\cal T}^{n_1,\mu_1}({\vec p}_1)} g^{\ka}(T_1)\,
{\cal P}_{r_1,n_1}\bigl(\log
(\sup({|\vec{p_1}| \over \ka},
{\ka \over m}))\bigl)
\label{snd}
\eqe
\[
\times
\sup_{T_2\in {\cal T}^{n_2,\mu_2}({\vec p}_2)} g^{\ka}(T_2)\,
{\cal P}_{r_2,n_2}\bigl(\log (\sup({|\vec{p_2}| \over \ka},
{\ka \over m}))\bigl)\ ,
\]
where we used the notations of (\ref{fequ})
and $\,\vec p_1:=\,(p_1,\ldots,p_{n_1-1},p')\,$,
$\,\vec p_2:=\,(-p',p_{n_1},\ldots,p_{n})\,$.
We pick two trees  $T_{1,\max}^{\ka}\,$ and $T_{2,\max}^{\ka}\,$,
which realize the $\sup$'s in (\ref{snd}) and
define the tree $T^{\ka}\,$ to be given by 
$T_{1,\max}^{\ka}\cup T_{2,\max}^{\ka}\cup \ell'$, 
where $\ell'\,$ is the internal line of the new tree  
$T\,$ joining $T_{1,\max}^{\ka}\,$ and $T_{2,\max}^{\ka}\,$.
This line carries the momentum $-p'$
(cf. (\ref{fequ})). We  attach the weight 2 to $\ell'\,$. 
We obviously have $T^{\ka}\in
{\cal T}^{n,\mu}\,$. Therefore integrating (\ref{snd}) from $\ka\,$ to
$\Lao+m\,$ (using again (\ref{bd})) the result is bounded by 
\[
g^{\ka}(T^{\ka})\ {\cal P}_{r_1,n_1}\bigl(\log
(\sup({|\vec{p_1}| \over \ka},
{\ka \over m}))\bigl)\,
{\cal P}_{r_2,n_2}\bigl(\log (\sup({|\vec{p_2}| \over \ka},
{\ka \over m}))\bigl)\ \log({|\vec p| \over \ka})
\]
\eq
\le \sup_{T\in{\cal
T}^{n,\mu}({\vecp})}\,g^{\ka}(T)\
{\cal P}_{r,n}\bigl(\log(\sup({|\vec{p}| \over \ka},
{\ka \over m}))\bigl)\ .
\label{sndd}
\eqe\qed

The present bounds seem close to optimal. 
They show for example that the high energy behaviour is not
deteriorated if only one single external momentum becomes small,
since our trees do not contain vertices of coordination number 2.
In particular for $n\,$ small (6,8,...) the number of weighted 
trees to be considered and thus the bound is easily explicited.
For $n=6$ we find three different trees\footnote{ When taking into
  account the Remark after Proposition 2, one finds that for 
 $n=6,\ r=2\,$ only the last of the 3 weight factors above appears,
a fact in accord with (trivial) direct calculation.
For $r\ge 3\,$ we again obtain all 3 types 
of trees.}, up to permutations of the external
momenta. Their weight factors $g^m(T)\,$ are
$\bigl(\sup(m,\, |p_1+p_2|)\; \sup(m,\, |p_1+p_2+p_3|)\bigl)^{-1}\;$,\\
$\bigl(\sup(m,\, |p_1+p_2|)\;\sup(m,\,  |p_3+p_4|)\bigl)^{-1}\;$,
$\bigl(\sup(m,\, |p_1+p_2+p_3|)\bigl)^{-2}\;$. From them, from 
the geometry of the external momenta and from
Proposition 2,  we read off the bound on the six point function.
\noindent
%%%%%%%%%%%%%%%%%%%%%%%%%%%%%%%%%%%%%%%%%%%%%%%%%%%%%%%%%%%%%%%

\end{document}